# Variation of photoluminescence spectral line shape of monolayer WS$_2$

*Yongjae Kwon,*[a] *Kangwon Kim,*[a] *Wontaek Kim,*[b] *Sunmin Ryu,*[b,c] *Hyeonsik Cheong*[a,*]

[a] Department of Physics, Sogang University, Seoul 04107, Korea

[b] Department of Chemistry, Pohang University of Science and Technology (POSTECH), Pohang 37673, Korea

[c] Division of Advanced Materials Science, Pohang University of Science and Technology (POSTECH), Pohang 37673, Korea

[*]E-mail: hcheong@sogang.ac.kr

## Abstract

The origin of the variation of photoluminescence (PL) spectra of monolayer tungsten disulfide (WS$_2$) is investigated systematically. Dependence of the PL spectrum on the excitation power show that the relatively sharp component corresponds to excitons whereas the broader component at slightly lower energy corresponds to negatively charged trions. PL imaging and second harmonic generation measurements show that the trion signals are suppressed more than the exciton signals near the edges, thereby relatively enhancing the excitonic feature in the PL spectrum and that such relative enhancement of the exciton signals is more pronounced near approximately armchair edges. This effect is interpreted in terms of depletion of free electrons near the edges caused by structural defects and adsorption of electron acceptors such as oxygen atoms.

**Keywords**

Tungsten disulfide, Photoluminescence, Exciton, Trion, Second harmonic generation

## 1. Introduction

Two-dimensional (2D) layered transition-metal dichalcogenides (TMDs) have attracted much interest recently owing to remarkable 2D physical properties that may be useful for novel device applications. For example, field effect transistors based on TMDs have shown a high on-off ratio [1]. Also, strong luminescence from many monolayer TMDs due to a direct band gap and a large exciton binding energy [2-6] may also be advantageous for nanoscale optoelectronic devices such as light-emitting devices [7, 8], photodetectors [9], or solar cells [10]. In addition, an optical logic gate by selective control of light helicity [11] has been demonstrated by utilizing the valley degree of freedom in TMDs [12-15]. Tungsten disulfide ($WS_2$) is one of the most studied TMDs and is frequently used to study valley effects or to make heterostructures with other TMDs.

The optical properties of TMDs are often dominated by the excitonic features because of large exciton binding energies owing to strong Coulomb interaction due to reduced screening in 2D materials [2-5]. Thanks to the strong Coulomb interaction, trions, a bound state of an exciton and an electron or a hole, are often observed with binding energies of tens of meV's [2, 16]. Because of the excitonic nature of the photoluminescence (PL), PL characterizations are used routinely to identify the number of layers or to estimate the quality of TMDs samples, etc. However, the reported PL spectra of monolayer $WS_2$ sample show a large variation [17, 18] unlike the case of $MoS_2$. At room temperature, monolayer $WS_2$ usually have strong PL signals at ~2.0 eV. Whereas some samples show a broad PL peak, others have a sharp peak at a slightly higher energy. In this paper, we try to clarify the origin of the variations by analyzing the excitation power dependence and spatial variation of the PL spectra of monolayer $WS_2$ samples.

## 2. Materials and methods

Monolayer $WS_2$ samples were prepared by mechanically exfoliating single crystal $WS_2$ flakes (HQ Graphene) onto Si substrates with a 285-nm $SiO_2$ layer. The number of layer was confirmed by combination of optical contrast and Raman spectroscopy. Raman and PL measurements were carried out using a home-built confocal microscope system using the 488-nm (2.54 eV) line of an $Ar^+$ laser and the 514.4-nm (2.41 eV) line of a diode-pumped-solid-state (DPSS) laser, respectively. The laser beam was focused onto a sample by a 50× objective lens (0.8 N.A.). The scattered light was collected and collimated by the same objective lens, dispersed by a Jobin-Yvon Horiba iHR550 spectrometer (2400 grooves/mm for Raman and 300 grooves/mm for PL), and detected with a liquid-nitrogen-cooled back-illuminated charge-coupled-device (CCD) detector. The laser power was kept at 50 μW to avoid damaging the sample, except for the excitation power dependence measurements. A microscope-based polarized second harmonic generation (SHG) system, described elsewhere in detail [19], was used to determine the crystallographic orientation of the sample. Briefly, a 140-fs Ti:sapphire laser (Coherent, Chameleon) with a wavelength of 820 nm and a repetition rate of 80 MHz was focused onto a 1.0-μm spot with a 40× objective. The SHG signal centered at 410 nm was collected by same objective lens and guided to a spectrograph (Andor, Shamrock 303i) equipped with a thermoelectrically cooled CCD (Andor, Newton). The angle between the $WS_2$ lattice and the laser polarization was varied by rotating the sample on a rotation stage with accuracy better than 0.2°. Parallel or perpendicular polarization of SHG signal was selected by an analyzer in front of the spectrograph.

# 3. Results and discussion

## 3.1 Variation of PL spectra from monolayer WS$_2$ samples

Figures 1(a) and (b) show optical images and Raman spectra of three monolayer WS$_2$ samples. The E′ and A$_1$′ modes, which correspond to the E$_{2g}^1$ and A$_{1g}$ modes due to intra-layer vibrations along the in-plane and out-of-plane directions in bulk WS$_2$, are observed at ~356 and ~417 cm$^{-1}$, respectively [20, 21]. The separation between the two peaks (~61 cm$^{-1}$) is consistent with the literature value for monolayer WS$_2$ [20-22]. Figure 1(c) shows PL spectra of the three samples. There is a large variation in the peak position and the line shape. Residual strain or variations in doping density influence the PL spectrum of TMDs materials [23, 24]. For example, the A$_1$′ mode, which is sensitive to doping [23], is slightly red-shifted in the Raman spectrum of sample 1 with respect to those in the other two spectra, which indicates that there is some doping. The different line shapes also indicate that the contributions of different components of the PL vary among samples. In order to find the origin of the PL spectrum variations, we studied sample 3 in more detail.

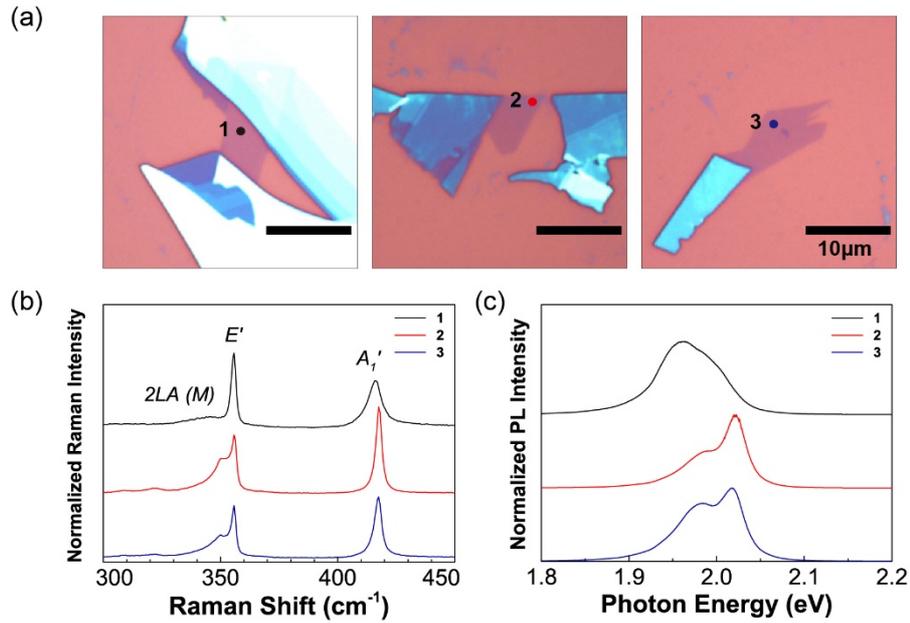

**Fig. 1.** (a) Optical microscope images of monolayer WS$_2$ samples. Normalized (b) Raman and (c) PL spectra of monolayer WS$_2$ for the three samples in (a).

### 3.2 Excitation power dependence of PL

Figure 2(a) shows the dependence of the PL spectrum from sample 3 on the excitation laser power from 5 to 200 μW. At low excitation powers, the spectrum is dominated by a relatively sharp peak at ~2.02 eV. A much weaker signal is seen at a lower energy of ~1.98 eV. As the excitation power increased, the lower energy peak grows much faster than the higher energy peak. The PL spectra were deconvoluted using a double-Lorentzian function, and the results are summarized in Figs. 2(b) and (c). The intensity reversal and the peak position shift are consistent with reported behaviors of exciton and trion peaks from WS$_2$. The PL signal from trions, bound states of two electrons and a hole or one electron and two holes, usually increase more rapidly than the exciton PL [25, 26]. Furthermore, the trion states tend to redshift at higher excitation power due to the shift of the Fermi level [25-28]. Because WS$_2$ samples without intentional doping are n-type [29], the trions are negatively charged. Therefore, we assign the higher energy peak to the exciton (X) recombination and the lower energy one to the negatively charged trion (X⁻) recombination.

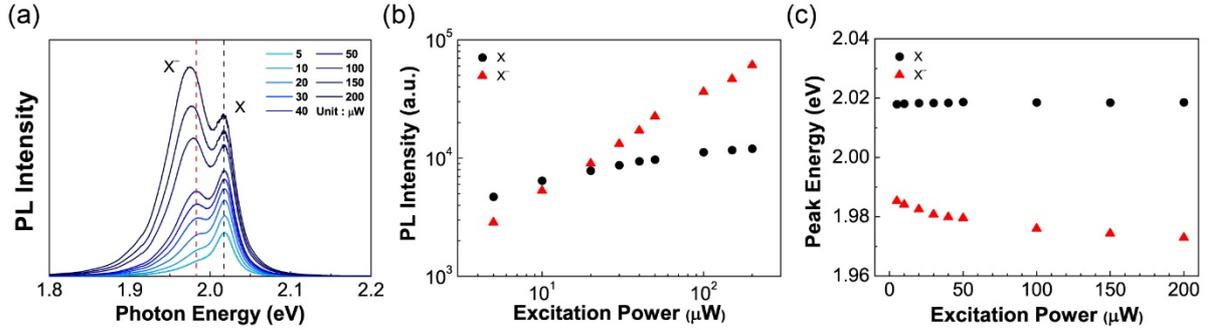

**Fig. 2.** (a) Excitation power dependence of PL spectrum of sample 3. (b) Intensities and (c) peak energies of exciton (X, black circle) and trion (X⁻, red triangle) signals as a function of excitation power, respectively.

### 3.3 Spatial variation of PL

In order to study the spatial variation of the PL spectrum from monolayer $WS_2$, we carried out PL imaging of sample 3. Figure 3(a) show PL spectra taken from the area marked by a white-dashed box in Fig. 3(b) in 1-μm steps. Results of deconvolution into double Lorentzian are also shown. The relative intensities of exciton and trion signals vary: trion seems to be stronger in the middle of the sample. For better comparison, we carried out PL imaging of the area in Fig. 3(b) in 0.5-μm steps and deconvoluted the exciton and trion components from each spectrum. Figures 3(c) and (d) show the intensity maps of exciton and trion PL signals, respectively, and Fig. 3(e) is the ratio of the former to the latter. Although the intensity is stronger in the middle part of the sample for both exciton and trion signals, the exciton signal is relatively stronger near the edges, which is what we see in Fig. 3(a). Due to structural defects near the edges, it is reasonable to expect that the PL signal is reduced near the edges. In order to see if the enhancement of the exciton to trion ratio at the edges has any correlation with the orientation of the edges, we carried out second harmonic generation (SHG) measurements. Since the intensity of the SHG signal is strongly sensitive to the relative angle between the principal axis of the crystal and the polarization of the incident light [30-33], the crystallographic orientation of the sample can be determined from polarized SHG measurements. Figure 4(b) shows the polarization dependence of the SHG signal with the relative angle $\theta - \theta_0$ defined in Fig. 4(c). The maximum signal in parallel polarization (minimum in cross polarization) corresponds to the armchair direction [30-33]. Figure 4(c) is an overlay of the determined crystallographic directions with Fig. 3(e). It appears that the relative enhancement of the exciton signal is stronger for approximately armchair edges. Since free electrons in addition to excitons are required to form negatively charged trions, the enhanced exciton signal or equivalently, suppression of the trion signal is an indication of reduced electron density near the edges. This is reasonable because presence of trap states due to a large density of structural defects near the edges would inevitably reduce the density of free carriers. Oxygen chemisorption or physisorption is often cited as the main mechanism for reduced electron density as oxygen atoms act as acceptors [34-40]. The observation that the effect seems to be stronger near approximately armchair edges imply that such chemisorption or physisorption is more effective near such edges. Although there is no systematic studies yet on the different sensitivity to chemisorption or physisorption at different edge types of $WS_2$, we would like to point to similar work on graphene in which it was found that the armchair edges tend to be more chemically active [41, 42].

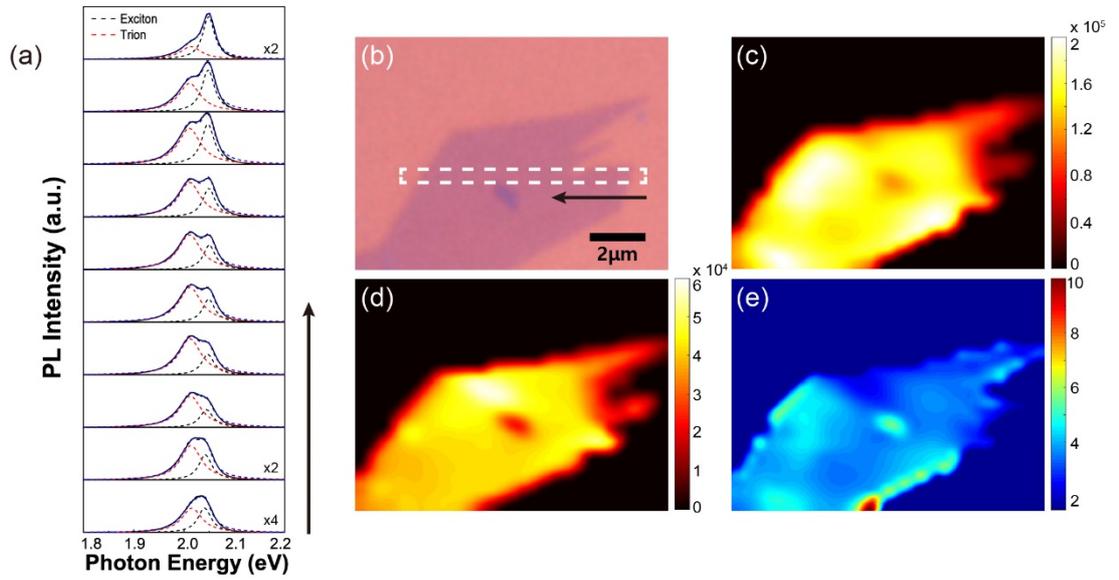

**Fig. 3.** (a) Position dependence of PL spectrum measured in the white box area of monolayer WS$_2$ shown in the optical microscope image of (b). The scan directions are indicated by black arrows in (a) and (b). Intensity maps of (c) exciton and (d) trion signals. (e) Intensity ratio image of exciton to trion.

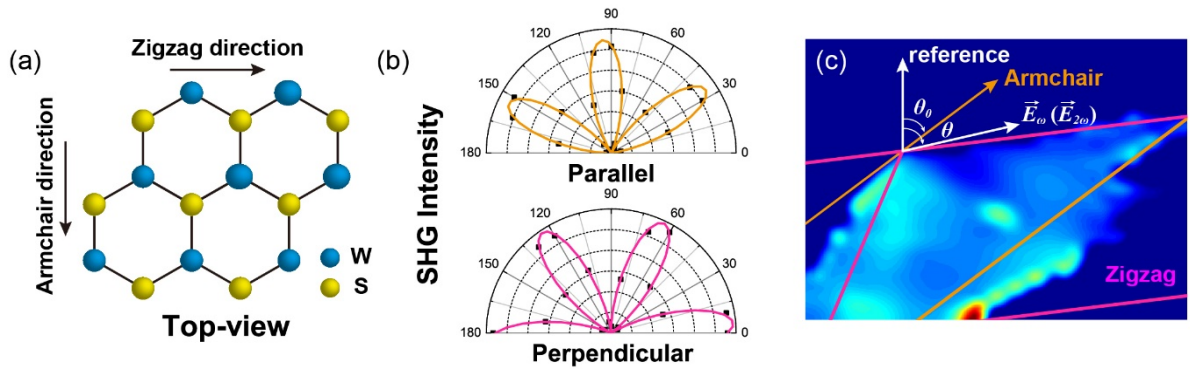

**Fig. 4.** (a) Crystal structure of WS$_2$ (top view). (b) Polar plots of SHG intensity in parallel and perpendicular polarizations as a function of angle $\theta$ relative to the laboratory reference direction. $\theta_0$ is the offset angle between the armchair direction and the reference direction as shown in (c). (c) Intensity ratio map of Fig. 3(e) overlaid with relevant crystallographic directions.

## 4. Conclusions

PL measurements have shown that the PL spectrum of monolayer $WS_2$ is a convolution of signals from excitons and trions, and the ratio between them varies in different samples or in different areas of the same sample. From PL imaging measurements, we found that the trion signal is more suppressed than the exciton signal near edges of monolayer $WS_2$, resulting in spectra with enhanced exciton signals. Polarized SHG measurements show that such relative enhancement of the exciton signal is more pronounced near approximately armchair edges. The suppression of trions is attributed to reduction of free electron density due to structural defects near edges through chemisorption or physisorption of adsorbates such as oxygen atoms.

## Acknowledgements

This work was supported by the National Research Foundation (NRF) grant funded by the Korean government (MSIT) (NRF-2016R1A2B3008363 and 2017R1A5A1014862, SRC program: vdWMRC center) and by a grant (No. 2011-0031630) from the Center for Advanced Soft Electronics under the Global Frontier Research Program of MSIT.

The authors declare no competing financial interest.